\documentclass{ws-ijmpa}
\begin{document}

\markboth{G.~L.~Klimchitskaya, M.~Bordag \& V.~M.~Mostepanenko}
{Comparison Between Experiment and Theory for the Thermal Casimir Force}

%
\catchline{}{}{}{}{}
%

\title{COMPARISON BETWEEN EXPERIMENT AND THEORY FOR THE
THERMAL CASIMIR FORCE}

\author{
G.~L.~KLIMCHITSKAYA
}

\address{North-West Technical University,
 Millionnaya Street 5, St.Petersburg,
191065, Russia{\protect \\}
Galina.Klimchitskaya@itp.uni-leipzig.de
}

\author{
M.~BORDAG
}

\address{Institute for Theoretical
Physics, Leipzig University, Postfach 100920,
D-04009, Leipzig, Germany
}
\author{
V.~M.~MOSTEPANENKO
}

\address{Noncommercial Partnership
``Scientific Instruments'',
Tverskaya Street 11, Moscow,
103905, Russia
}

\maketitle

\begin{history}
\received{17 November 2011}
\revised{14 January 2012}
\end{history}

\begin{abstract}
We analyze recent experiments on measuring the thermal
Casimir force with account of possible background
effects. Special attention is paid to the validity of the
proximity force approximation (PFA) used in the comparison
between the experimental data and computational results in
experiments employing a sphere-plate geometry.
The PFA results are compared
with the exact results where they are available.
The possibility to use fitting procedures in theory-experiment
comparison is discussed. On this basis we reconsider
experiments exploiting spherical
lenses of centimeter-size radii.

\end{abstract}

\ccode{PACS numbers: 12.20.-m, 12.20.Fv, 78.20.Ci}

\section{Introduction}

During the last ten years a lot of attention was devoted to
measurements of the Casimir force\cite{1} and to comparison between the
experimental data and theoretical predictions.\cite{2}\cdash\cite{8}
It is well known that predictions of the Lifshitz theory for the
thermal Casimir force strongly depend on models of the dielectric
permittivity used in computations. Typically the dielectric permittivity
of ideal dielectric materials (i.e.,
isolators of infinitely high resistivity) is
determined by the core electrons and can be presented in an oscillator form
\begin{equation}
\varepsilon_c(\omega)=1+\sum_{j=1}^{K}
\frac{g_j}{\omega_j^2-\omega^2-i\gamma_j\omega},
\label{eq1}
\end{equation}
\noindent
where $K$ is the number of oscillators, $\omega_j$ are the oscillator
frequencies, $g_j$ are the oscillator strengths, and $\gamma_j$ are
the damping parameters.
Real dielectrics, however, at any nonzero temperature possess some
static dc conductivity $\sigma_0(T)$. As a result, their
dielectric permittivity is
given by
\begin{equation}
\varepsilon_d(\omega)=\varepsilon_c(\omega)+
i\frac{4\pi\sigma_0(T)}{\omega}.
\label{eq2}
\end{equation}
\noindent

For real metals rather good representation for the
dielectric permittivity is given by
the Drude model
\begin{equation}
\varepsilon_D(\omega)=\varepsilon_c(\omega)-
\frac{\omega_p^2}{\omega[\omega+i\gamma(T)]},
\label{eq3}
\end{equation}
\noindent
where $\omega_p$ is the plasma frequency and $\gamma(T)$ is the relaxation
parameter. In the region of infrared frequencies it holds
$\gamma(T)\ll\omega$ and (\ref{eq3}) converts to the so-called plasma
model
\begin{equation}
\varepsilon_p(\omega)=\varepsilon_c(\omega)-
\frac{\omega_p^2}{\omega^2}.
\label{eq4}
\end{equation}
\noindent
This model disregards relaxation which does not play any role at so
high frequencies. It was proved\cite{5,6,9,10} that the Lifshitz theory
combined with the most realistic permittivities (\ref{eq2}) and (\ref{eq3})
violates the third law of thermodynamics (the Nernst heat theorem) for
dielectrics and metals with perfect crystal lattices, respectively.
This result was originally proven for the configuration of two
plane-parallel plates. In Ref.~\refcite{10a} the violation of the
Nernst heat theorem for metals with perfect crystal lattices described
by the Drude model and for dielectrics with included dc conductivity
was demonstrated for the configuration of a sphere above a plate.
This makes particularly important the experimental confirmation of one
or other model of dielectric permittivity mentioned above.

In this paper we analyze several recent experiments which are sufficiently
exact to discriminate between theoretical results computed using
different models for the dielectric permittivity. We discuss the validity
of the approximations employed to compare experiment with theory and the
role of some background effects, such as patch potentials. Section~2
is devoted to the accuracy of the proximity force approximation (PFA).
In Sec.~3 we consider the experimental results\cite{11,12} obtained by means
of a micromachined oscillator which are consistent with (\ref{eq4}) but
exclude (\ref{eq3}). Here we also discuss the possibility of using the
fitting procedures in theory-experiment comparison and the recently
proposed new model of patch potentials.\cite{13}
Section~4 contains new information obtained from experiments with
semiconductor\cite{14}\cdash\cite{16} and dielectric\cite{17,18} test
bodies. These experiments are consistent with the dielectric permittivity
(\ref{eq1}) but exclude (\ref{eq2}). The critical analysis of
claimed observation of the thermal Casimir force\cite{19} is
presented in Sec.~5. Here we show that the long-separation data of this
experiment better agrees not with (\ref{eq3}), as claimed, but with the
dielectric permittivity (\ref{eq4}).
In Sec.~6 the reader will find our conclusions.

\section{Is the PFA Exact Enough for Theory-Experiment Comparison?}

All precise measurements of the Casimir force performed to date use
the configuration of a sphere (or a spherical lens) above a plate.
To calculate the theoretical Casimir force $F$ in sphere-plate geometry
one first calculates the free energy ${\cal F}$ between two parallel
plates and then uses the PFA in the form\cite{5}
\begin{equation}
F(d,T)=2\pi R{\cal F}(d,T),
\label{eq5}
\end{equation}
\noindent
where $d$ is the shortest separation between the  sphere and the
plate and between two plates, $R$ is the radius of the sphere.
The error introduced from the use of the PFA can be estimated from
the comparison with exact results for sphere-plate geometry.
Recently it was shown\cite{20,21} that at $T=0$ the first two leading
terms for the electromagnetic Casimir energy in the
configuration of an ideal metal sphere above
an ideal metal plate are given by
\begin{equation}
E(d)=-\frac{\pi^3R\hbar c}{720d^2}\left[1+\left(\frac{1}{3}-
\frac{20}{\pi^2}\right)\frac{d}{R}\right]=
-\frac{\pi^3R\hbar c}{720d^2}\left(1-1.69\frac{d}{R}\right).
\label{eq5a}
\end{equation}
\noindent
For the force between a sphere and a plate
from (\ref{eq5a}) one finds
\begin{equation}
F(d)=-\frac{\pi^3R\hbar c}{360d^3}\left[1+\left(\frac{1}{6}-
\frac{10}{\pi^2}\right)\frac{d}{R}\right]=
-\frac{\pi^3R\hbar c}{360d^3}\left(1-0.85\frac{d}{R}\right).
\label{eq5b}
\end{equation}
\noindent
Thus, for a force, the leading correction to the PFA result is
equal to $-0.85d/R$.
This is really a small correction if to take into consideration
that the experimental range of parameters is given by
\begin{equation}
0.001\leq\frac{d}{R}\leq 0.007.
\label{eq6}
\end{equation}

For an ideal metal sphere above an ideal metal plane it was also
proven\cite{22} that the PFA gives the same values of the Casimir
force at $T=0$ and of the thermal correction to it as does the
exact theory in the zeroth order of $d/R$ [with exception of only
extremely low $T\ll\hbar c/(2k_BR)$]. Thus, for ideal metals
the use of the PFA in sphere-plate geometry may result in only
small errors less than $d/R$.

For a sphere and a plate made of metals described by the Drude and
plasma models the exact computations of the Casimir force were
performed\cite{23,24} for $d/R>0.1$. Deviations between the
predictions of the PFA and the exact theory were described in terms of
the quantity $q=F_p/F_D$, where $F_p$ and $F_D$ are the Casimir forces
calculated using the simple plasma  and Drude models, respectively,
i.e., using Eqs.~(\ref{eq3}) and (\ref{eq4}) with
$\varepsilon_c(\omega)=1$. It was shown that deviations between the
values of $q$ computed using the exact theory and the PFA decrease
from 9.2\% to 2.5\% when $d/R$ decreases from 5 to 0.1.

In the framework of the PFA in the high-temperature regime
$T\gg  T_{\rm eff}=\hbar c/(2dk_B)$ or, alternatively, at
large separations $d\gg \hbar c/(2k_BT)$ it holds $q=2$
because under these conditions ${\cal F}_p(d,T)=2{\cal F}_D(d,T)$.
It was claimed, however, that at large separations the exact
theory leads to $q=3/2$ instead of a factor $q=2$ as follows from
the PFA.\cite{23,24}
This statement formulated in so general form
is somewhat misleading because it does not take into
consideration the application region of the PFA. In fact the value
of $q$ depends on the interplay between different parameters.
Thus, it  really holds that
\begin{equation}
q=\frac{3}{2}\quad\mbox{for}\quad
\frac{2\pi c}{\omega_p}\ll R\ll d.
\label{eq7}
\end{equation}
\noindent
This is the limit of extremely large separations (much
larger than the sphere radius). The condition (\ref{eq7})
is outside the experimental region (\ref{eq6}) and outside
the application region of the PFA. Because of this it is
not reasonable to compare the result $q=3/2$ with $q=2$,
as obtained from the PFA.

There is also another case of large separations outside the
application region of the PFA where one has\cite{25}
\begin{equation}
q=1\quad\mbox{for}\quad
d\gg R,{\ \ }d\gg \frac{\hbar c}{2k_BT},
{\ \ }R\leq\frac{2\pi c}{\omega_p}.
\label{eq8}
\end{equation}

Finally, exact computations lead\cite{26} to the same result as
the PFA
\begin{equation}
q=2\quad\mbox{for}\quad
\frac{\hbar c}{2k_BT}\ll d\ll R.
\label{eq9}
\end{equation}
\noindent
This is just the region of large separations which simultaneously
belongs to the application region of the PFA. For instance,
exact computations show\cite{23,24} that at $d=5\,\mu$m $q$
increases from approximately 1.48 to 1.63 when $d/R$ decreases from
2.5 to 0.5. With further decreasing of $d/R$ the quantity $q$
goes to 2, as follows from the PFA.

One can conclude that in the region of the experimental parameters
(\ref{eq6}) the PFA is well applicable for the comparison between
experiment and theory in the configuration of a perfectly shaped
sphere (spherical lens) above a plane plate.

\section{Experiments Between Metallic Test Bodies Using a
Micromachined Oscillator}

In several successive experiments\cite{11,12,27,28} the Casimir pressure
between two Au plates was determined from dynamic measurements of the
gradient of the Casimir force betwen a sphere and a plate. These
experiments were performed in the separation region
$160\,\mbox{nm}\leq d\leq 750\,$nm. Their results gave rise to
continuing hot discussions because contrary to expectations they were
found to exclude the dielectric permittivity of the Drude model and to
be consistent with the dielectric permittivity of the plasma model.
Here we present the comparison of the measurement data with theory\cite{11,12}
in the region of the largest separations $700\,\mbox{nm}\leq d\leq 750\,$nm
overlapping with another recent experiment\cite{19} leading to the opposite
conclusions (see Sec.~5).

\begin{figure*}[t]
\vspace*{-7.8cm}
\centerline{\hspace*{4.cm}\psfig{file=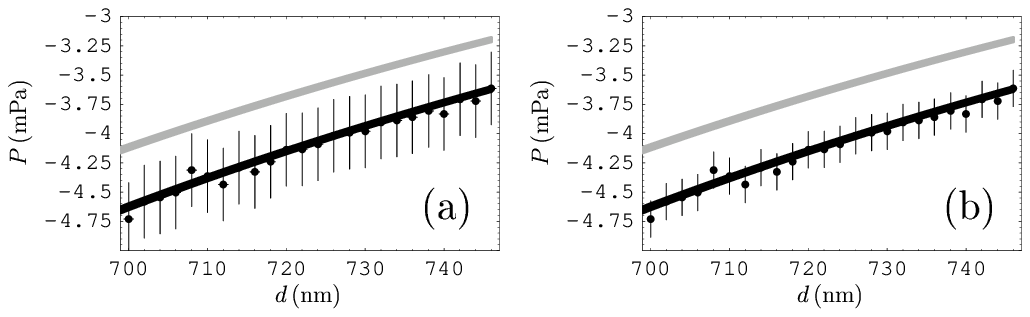,width=24cm}}
\vspace*{-22.4cm}
\caption{The experimental data for the Casimir
pressure between two parallel plates measured
by means of micromechanical torsional oscillator as a function
of separation are shown as crosses.
The arms of the crosses indicate the total experimental
errors  determined at
(a) 95\%  and (b) 67\% confidence level.
The grey and black lines show
the theoretical Casimir pressures computed
 using the Drude and plasma model approaches,
 respectively.
 The thickness of the lines indicates the total
 theoretical errors.}
\end{figure*}
In Fig.~1(a,b) the mean measured Casimir presures are shown as
crosses. The arms of the crosses indicate the total experimental
errors in separations and pressures determined (a) at a 95\%
confidence level and (b) at a 67\% confidence level.
The upper (grey) and lower (black) lines show theoretical Casimir
pressures computed using the Drude and plasma model approaches.
Note that the Drude model approach does not reduce to the use of
dielectric permittivity (\ref{eq3}). It exploits the tabulated optical
data\cite{29} extrapolated to lower frequencies by means of the
simple Drude model. The point is that (\ref{eq3}) does not provide
a sufficiently accurate picture for the optical data of Au in the
frequency region below the first absorbtion band. To model the
optical data in that region using Eq.~(\ref{eq3}) one should
consider\cite{30} the frequency-dependent relaxation parameter
$\gamma(\omega,T)$. The experimental results\cite{11,12} exclude
the predictions of the Drude model approach over the entire
measurement range from 162 to 746\,nm. Figure~1(a,b) convincingly
demonstrates this fact at $d>700\,$nm.
Possible theoretical reasons why the experimental data\cite{11,12}
are consistent with the plasma model, which does not include
dissipation, and exclude a more correct Drude model are discussed
in Ref.~\refcite{30a}.

\subsection{What is changed if one uses the fitting procedure
between the data and the theoretical Casimir force}

The comparison between experiment and theory shown in Fig.~1(a,b) is
independent in the sense that the measured Casimir forces were not used
to specify any theoretical parameter. Let us now admit that some
parameter of the theoretical Casimir force is unknown. The question
arises whether it is possible to determine it by fitting the measurement
data to theory and simultaneously choose between two different
theoretical approaches. To answer this question, we perform the
$\chi^2$-fit of the data in Fig.~1(a,b) to the perturbation
expansions of the Lifshitz formula for the Casimir pressure using the
simple plasma and Drude models.\cite{5} Such expansions are well
applicable at $d>700\,$nm. Thus, for the plasma model one has\cite{5}
\begin{equation}
P_p^{\rm th}(d,\delta)=-\frac{\pi^2\hbar c}{240d^4}\left(1-
\frac{16}{3}\,\frac{\delta}{d}+24\frac{\delta^2}{d^2}\right),
\label{eq10}
\end{equation}
\noindent
where $\delta=\lambda_p/(2\pi)=c/\omega_p$ is the penetration depth
of electromagnetic fluctuations into the metal.

We now assume for a moment that the value of $\delta$ for Au is not known
and find it from the best fit between $P_p^{\rm th}$ and 24 experimental
data points $P_i$ in Fig.~1(b). The minimization of the
$\chi^2$-function
\begin{equation}
\chi^2=\sum_{i-1}^{N}
\frac{|P_i-P^{\rm th}(d_i,\delta_1,\,\ldots,\,\delta_k)|}{(\Delta P_i)^2},
\label{eq11}
\end{equation}
\noindent
where $N=24$, $\Delta P_i$ are the total experimental errors determined
at a 67\% confidence level, $P^{\rm th}=P_p^{\rm th}$,
$\delta_1=\delta$ and $k=1$ leads to
$\chi_{\min}^2=1.88$ and $\delta=21.25\,$nm. This result is in a very good
agreement with the independently known value $\delta^{\rm (Au)}=22\,$nm.
The value of the $\chi^2$-probability
$P(\chi^2>\chi_{\min}^2)$, i.e., the probability of an event that in a new
measurement the value of $\chi^2$ larger than $\chi_{\min}^2$ will be
obtained, can be found using the number of degrees of freedom $f=N-k=23$.
The result is $P(\chi^2>\chi_{\min}^2)=0.9999$ which confirms that
the plasma model is in a very good agreement with the data.
This conclusion was obtained from the fit to the mean measured Casimir
pressures. The same fit repeated with each of 33 sets of individual
measurements results in
\begin{equation}
0.07\leq P(\chi^2>\chi_{\min}^2)\leq 0.9,\quad
17\,\mbox{nm}\leq \delta\leq 24\,\mbox{nm}.
\label{eq12}
\end{equation}

Now we repeat the fitting procedure but use the theoretical
pressures obtained from the Drude model approach\cite{5}
\begin{equation}
P_D^{\rm Th}(d,\delta)=-\frac{\pi^2\hbar c}{240d^4}\left(1-
\frac{16}{3}\,\frac{\delta}{d}+24\frac{\delta^2}{d^2}\right)
+
\frac{k_BT\zeta(3)}{8\pi d^3}\left(1-
6\frac{\delta}{d}+24\frac{\delta^2}{d^2}\right).
\label{eq13}
\end{equation}
\noindent
{}From the mean measured Casimir pressures in Fig.~1(b) one obtains
\begin{equation}
\chi_{\min}^2=2.42,\quad
P(\chi^2>\chi_{\min}^2)=0.9999, \quad
\delta=4.2\,\mbox{nm}.
\label{eq14}
\end{equation}
\noindent
This could be also considered as a very good agreement with the data
if the correct value of $\delta^{\rm (Au)}=22\,$nm were not
independently known.
The fit to different individual sets of measurements
leads to
\begin{equation}
0.07\leq P(\chi^2>\chi_{\min}^2)\leq 0.9,\quad
2.5\,\mbox{nm}\leq \delta\leq 6\,\mbox{nm}.
\label{eq15}
\end{equation}
\noindent
Keeping in mind that all the values of $\delta$ obtained from the fit
to the Drude model Casimir pressure (\ref{eq13}) are in complete
disagreement with the correct value, the theoretical description of the
data by means of Eq.~(\ref{eq13}) should be rejected in spite of a high
$\chi^2$-probability.

{}From the above one can conclude that the fit of the experimental data
to two competing theoretical approaches containing unknown parameters may
not allow for a reliable choice between these approaches.

\subsection{Could the patch effect compensate differences between
the experimental data and the Drude model prediction?}

It is common knowledge that the patch potentials due to grain structure of
Au coatings, surface contaminants etc. lead to additional electric force
which should be taken into account in precision measurements.
Using the earlier proposed model of patches\cite{31} it was shown\cite{28}
that they lead to a negligibly small contribution to the Casimir pressure.
Recently the so-called quasi-local model of patches was suggested\cite{13}
which leads to the orders of magnitude larger electric pressure than that
predicted in Ref.~\refcite{31}. In this connection it was claimed\cite{13}
that ``patches may render the experimental data at distances below
1 micrometer compartible with theoretical predictions based on the Drude
model''. In support of this claim a fit of the suggested patch model
to the difference between the measurement results\cite{11} and
the theoretical prediction for the Casimir pressure based on the Drude
model (called the residual signal) was performed. It was found that for
the maximum patch size approximately equal to $l_{\max}=1074\,$nm
and the root-mean-square voltage of $V_{rms}=12.9\,mV$ there is a
qualitative agreement between the residual signal and the fitted
patch pressure to within a few percent of the total measured pressure.
Such large patches cannot be connected with the grain structure of
the surface and were attributed to some hypothetical contaminants.\cite{13}

With respect to this result it should be stressed that the theoretical
pressures predicted by the Drude model were calculated\cite{13}
not as in Refs.~\refcite{11,12} (i.e., using the tabulated optical
data\cite{29} extrapolated to low frequencies using the simple Drude model)
but with the help of Eq.~(\ref{eq3}) with three oscillators.
As was noted above, this leads to
significant deviations in the dielectric permittivity and therefore
in the computational results for the Casimir pressure. Moreover, the
surface roughness was not taken into account in Ref.~13.
For comparison purposes in Fig.~2 we plot the magnitudes of the residual
signal as a function of separation for the Drude model theoretical
prediction as in Ref.~\refcite{13} (the upper set of crosses shown at
a 67\% confidence level) and as in Refs.~\refcite{11,12} (the lower set
of crosses shown at the same confidence level). As can be seen in Fig.~2,
at separation distances below $d=300\,$nm the residual signal
used in Ref.~\refcite{13}
deviates significantly from a more precise result.\cite{11,12}

It is worthy of note also that the agreement between the residual
signal and the patch pressure is up to a few percent of the total
measured pressure.\cite{13} This should be compared with the relative
differences between the theoretical predictions using the Drude and plasma
model approaches which are also of about a few percent of the total
measured pressures at separations below $1\,\mu$m.
Thus, the statement made reduces to saying
that the patch effect and the residual
signal are of the same order of magnitude.

To check whether the suggested patch model can be employed to explain
the residual signal $|P^{\rm expt}-P_D^{\rm th}|$, we have calculated
the value of $\chi_{\min}^2$ using Eq.~(\ref{eq11}) with
$P^{\rm th}=P_D^{\rm th}$, as computed in Ref.~\refcite{13}, and two
fitting parameters $\delta_1=l_{\max}$ and $\delta_2=V_{rms}$ specified
above. It was found that $\chi_{\min}^2\approx 700$. Taking into account
that the number of degrees of freedom is equal to $f=291$, from this we
obtain that $P(\chi^2>\chi_{\min}^2)=0$ with at least 8 zeros after a
decimal comma. This result means that the theoretical pressure-distance
dependence caused by the quasi-local patches\cite{13} is irrelevant to
the difference between the experimental data\cite{11,12} and the predictions
of the Lifshitz theory combined with the Drude model.

In spite of the above result obtained with respect to the quasi-local model
of patches, it would be interesting to perform direct measurements of
patch distributions on specially cleaned and prepared surfaces in high
vacuum, as are used in measurements of the Casimir force.
This would bring reliable model-independent information concerning the
role of patch potentials in Casimir physics.

\section{Experiments with Semiconductor and Dielectric Test Bodies}

Here we briefly review several recent experiments which demonstrate
that the Lifshitz theory with taken into account dc conductivity of
dielectric materials is excluded by the measurement data.

\subsection{Optical modulation of the Casimir force between an Au
sphere and Si plate}

\begin{figure*}[t]
\vspace*{-9.3cm}
\centerline{\hspace*{3.cm}\psfig{file=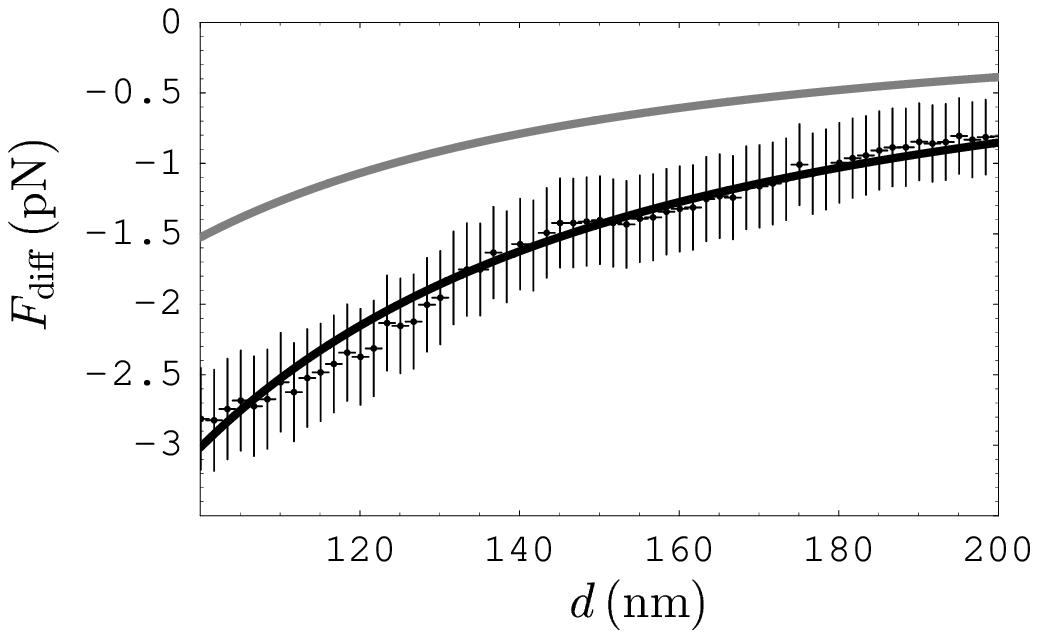,width=12cm}}
\vspace*{-4.cm}
\caption{ The experimental data for the difference Casimir
force between an Au sphere and Si plate measured
by means of AFM as a function of separation are shown as
crosses.
The arms of the crosses indicate the total experimental errors
determined at a 95\% confidence level.
The black and grey lines show
the theoretical difference forces computed
with neglected and included dc conductivity of Si plate
in the absence of light, respectively.
}
\end{figure*}
In this experiment\cite{14,15} an Au sphere was attached to the
cantilever of the atomic force microscope above a Si plate
illuminated with laser pulses. The measured quantity was the difference in
the Casimir forces between a sphere and a plate in the presence and in
the absence of laser light on the plate. In the absence of light Si was in
a dielectric state with the density of charge carriers of about
$n=5\times 10^{14}\,\mbox{cm}^{-3}$. In the presence of laser pulse the
density of charge carriers was higher up to 5 orders of magnitude.
Thus, in the presence of light Si was in metallic state with
$n>n_{cr}$, where $n_{cr}$ is the critical density of charge carriers
such that the phase transition from dielectric to metallic state occurs.

In Fig.~3 we demonstrate as crosses the typical measured differences
in the Casimir forces, $F_{\rm diff}$, versus separation between the
sphere and the plate.
The arms of the crosses indicate the total experimental errors determined
at a 95\% confidence level. The grey line shows the results of
computations using the Lifshitz theory with included contribution of free
charge carriers, both in the absence and in the presence of laser light
on a Si plate. As can be seen in Fig.~3, this theoretical approach is
excluded by the experimental data at a 95\% confidence level.
The black line shows the theoretical results obtained with included
contribution of free charge carriers in the presence of laser light
(i.e., when Si was in metallic state), but with charge carriers
disregarded in the absence of laser light on a Si plate (i.e., when Si
was in dielectric state).
{}From Fig.~3 it is seen that the black line is fully consistent with the
experimental data. Thus, the inclusion of dc conductivity of dielectric
Si in the Lifshitz theory is in contradiction to the optical
modulation experiment.

\subsection{The Casimir force between an Au sphere and an
indium tin oxide plate}

In this experiment\cite{16,16a} the atomic force microscope was used to measure
the Casimir force between an Au sphere and an indium tin oxide (ITO) plate
as a function of separation (for the details of calibration of a setup
see Ref.~\refcite{32}). Then the ITO plate was UV-treated and measurements
of the Casimir force were repeated. The experimental results for the
untreated ITO sample are shown as the lower set of crosses in Fig.~4(a).
These results are consistent with earlier measurements of the Casimir force
gradient between an Au sphere and an ITO plate.\cite{33,34}
The measurement data for the untreated sample are in good agreement
with theoretical results computed using the Lifshitz theory with
included contribution of free charge carriers (the lower pair of solid
lines forms the theoretical band; the thickness of this band is caused
by a freedom in the extrapolation of the measured optical data of ITO
to higher frequencies).
\begin{figure*}[t]
\vspace*{-7.5cm}
\centerline{\hspace*{3.cm}\psfig{file=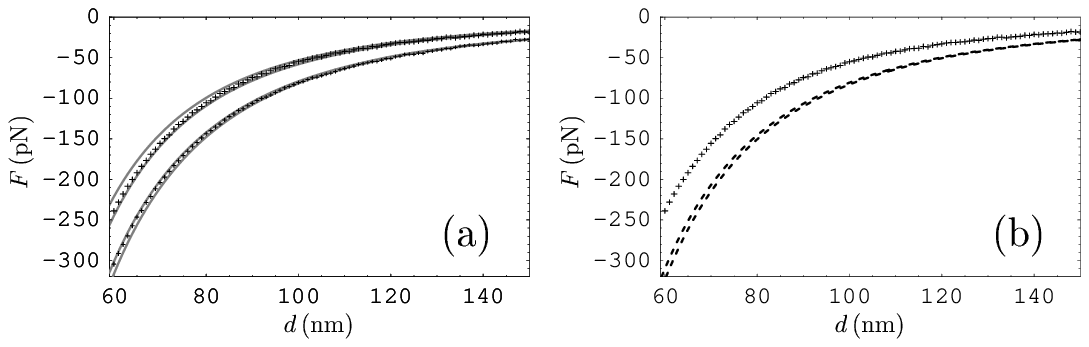,width=24cm}}
\vspace*{-22.4cm}
\caption{(a) The mean measured Casimir force as function of separation
is indicated as crosses
for the untreated (the lower set of crosses) and UV-treated (the upper
set of crosses) sample.
The theoretical Casimir force shown by the pairs of solid lines
  for the untreated sample is calculated with free charge carriers included
 and for the UV-treated sample with free charge carriers omitted.
(b) The mean measured Casimir force as a function of
separation (crosses) and the theoretical results computed with
included contribution of free charge carriers (two dashed lines)
for the UV-treated sample.
}
\end{figure*}

Quite unexpectedly, the measured Casimir force from the UV-treated
sample was found to differ significantly from the untreated one. It is
shown as the upper set of crosses in Fig.~4(a). In fact the decrease in the
force magnitude ranging
from 21\% to 35\% depending on separation was observed.
The same set of crosses
demonstrating the Casimir force from the UV-treated sample
is shown in Fig.~4(b).
The optical data for a UV-treated sample were measured and found almost
the same as for an untreated one.
This is in conflict with the fact that the magnitudes of the Casimir
force obtained after a UV treatment are much smaller than for an
untreated sample.
The computational results for the UV-treated sample with included
contribution of ITO charge carriers are shown by the pair of dashed
lines in Fig.~4(b). The computational results
 with disregarded contribution of ITO charge carriers are shown
by the pair of upper solid lines in Fig.~4(a). As can be seen in Fig.~4(a,b),
the inclusion of free charge carriers for the UV-treated ITO sample
results in complete disagreement between
the data and the theory, whereas the neglect of free charge carriers
makes theory consistent with the measurement data.
It was suggested\cite{16} that the UV treatment of ITO causes the phase
transition from metallic to dielectric state (this is confirmed by the
fact\cite{35} that the UV treatment leads to lower mobility of charge
carriers). If this is the case, than the neglect of free charge carriers
in the UV-treated ITO sample can be compared with the neglect of dc
conductivity in dielectric Si considered above.\cite{36}
It remains unexplained, however, why one should disregard the dc
conductivity of a dielectric body when calculating the Casimir force
in the framework of the Lifshitz theory.

\subsection{Frequency shift of center-of-mass oscillations
due to the Casimir-Polder force}

\begin{figure*}[t]
\vspace*{-9.cm}
\centerline{\hspace*{4.cm}\psfig{file=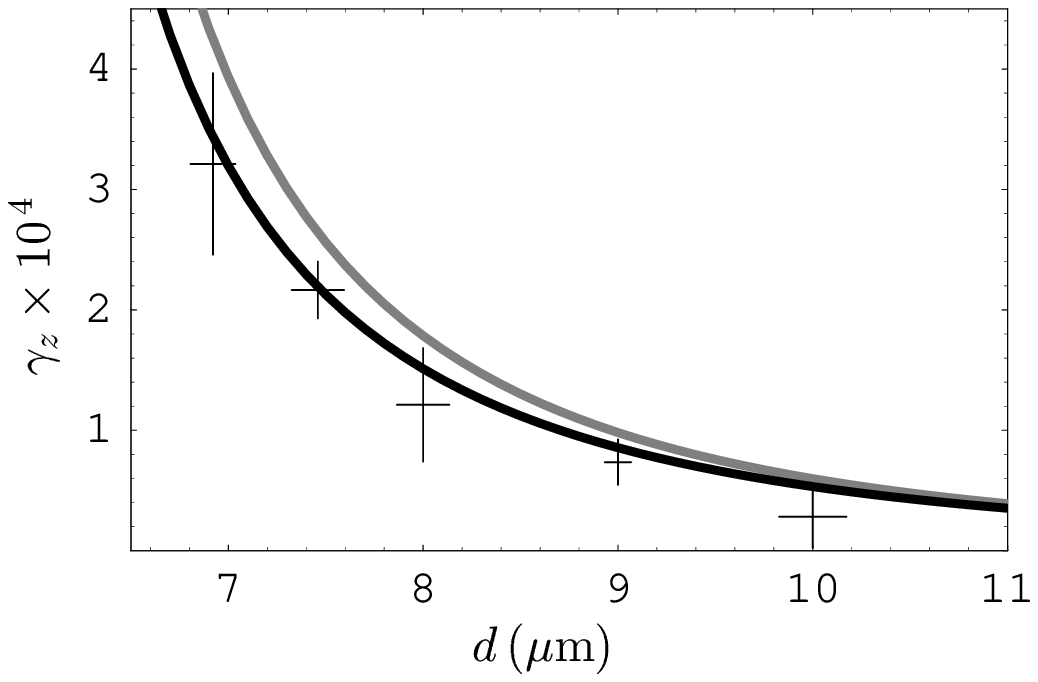,width=12cm}}
\vspace*{-4.cm}
\caption{The experimental data for the fractional change of the trap
frequency due to the Casimir-Polder force between ${}^{87}$Rb
atoms and fused silica plate are indicated by crosses
as a function of separation.
The total experimental errors are
determined at a 70\% confidence level.
The black and grey lines show the theoretical fractional
shift computed with neglected
and included dc conductivity of fused silica.
}
\end{figure*}
The Casimir-Polder force between an atom and a plate leads to a
fractional
shift $\gamma_z$ of the center-of-mass oscillation frequency $\omega_z$
of the Boze-Einstein condensate in the direction perpendicular to
the plate. This shift was measured\cite{17} for the condensate of
${}^{87}$Rb atoms and SiO${}_2$ dielectric plate at separations from
7 to $11\,\mu$m, i.e., in the region of thermal Casimir-Polder force,
both at thermal equilibrium and out of equilibrium when the temperature
of a plate was higher than that of environment.
The results obtained were compared with the Lifshitz theory and its
generalization for a nonequilibrium situation.
As an example, in Fig.~5 the experimental results for $\gamma_z$
as a function of separation are shown as crosses for the plate at $T=605\,$K
and an environment at $T=310\,$K.
The total experimental errors in both $\gamma_z$ and $d$ are indicated at
a 67\% confidence level. The black solid line shows the theoretical
results obtained\cite{17} with disregarded dc conductivity of SiO${}_2$.
These results are consistent with the measured data. The grey line in
Fig.~5 shows the theoretical results computed\cite{18} with
dc conductivity
of SiO${}_2$ included. As can be seen in Fig.~5, the theory with
taking into account dc
conductivity of dielectric SiO${}_2$ is excluded by the data.
This again confirms the phenomenological prescription\cite{5,6} that
in computations of the Casimir force the contribution of free charge
carriers of dielectrics and semiconductors of dielectric type should be
disregarded. Needless to say that such kind phenomenological
prescriptions cannot be considered as conclusive explanations and need
justification on the basis of first principles.

\section{Torsion Balance Experiments}

Contrary to the previous experiments which used spheres of
100--150\,$\mu$m radii or even atoms to measure the Casimir
(Casimir-Polder) force, torsion balance experiments exploit spherical
lenses with 15-20\,cm radii of curvature. The use of large lenses allows
to significantly increase the Casimir force but makes problematic
sufficiently precise characterization of the surface. Below we consider two
recent experiments employing large spherical lenses to measure the
Casimir force which arrive to the conflicting conclusions.

\subsection{Confirmation of the thermal Casimir force predicted by the
plasma model}

The first\cite{37} of two recent torsion balance experiments used a lens
of $R=20.7\,$cm radius of curvature and a plate both coated with Au.
The measured data for the force gradient over the separation region
from 0.48 to $6.5\,\mu$m were fitted to the sum of contributions
from the residual electric force and from the theory of the thermal
Casimir force based on the plasma model. The roughness corrections were
also taken into account. The fit used Eq.~(\ref{eq11}) with one fitting
parameter (the residual potential difference $V_{\rm res}$).

As a result, the minimum value of the $\chi^2$-function was found
$\chi_{\min}^2=513$ with $V_{\rm res}=20.0\pm 0.2\,$mV.
Taking into account that the number of degrees of freedom was $f=558$,
from this one obtains $P(\chi^2>\chi_{\min}^2)=0.91$.
This demonstrates the high level of agreement between the data and
the used theory of the Casimir force and supports the results of
Refs.~\refcite{11,12,27} and \refcite{28} considered in Sec.~3.

\subsection{Claimed observation of the thermal Casimir force predicted by the
Drude model}

The second\cite{19} of two recent torsion balance experiments used
a lens of $R=15.6\,$cm radius of curvature. Both the lens and the plate
were coated with Au. The measured data for the total force
$F_{\rm tot}(d)$
over the separation region from 0.7 to $7.3\,\mu$m were fitted to the
sum of the Casimir force, the hypothetical electrostatic force due to
large patches, and some constant contribution $-a$ called an offset
due to voltage offsets in the measurement electronics.\cite{19}

The Casimir force $F(d)$ was calculated in the framework of both the Drude
and the plasma  model approaches. The form of the force due to large
patches remained unknown. According to the authors,\cite{19}
``an independnet measurement of this electrostatic force with the required
accuracy is currently not feasible.'' It was suggested that there are
large patches of size $\lambda$ on Au-coated surfaces such that
\begin{equation}
d\ll\lambda\ll\sqrt{Rd}
\label{eq16}
\end{equation}
\noindent
due to absorbed impurities or oxides.
The electric force due to these patches was modeled by the term\cite{19}
\begin{equation}
F_{\rm patch}(d)=-\pi\epsilon_0 R\frac{V_{\rm rms}^2}{d},
\label{eq17}
\end{equation}
\noindent
where $\epsilon_0$ is the permittivity of the free space, and
$V_{\rm rms}$ is the magnitude of voltage fluctuations across the text
bodies. In view of this, after the application of voltage in order
to cancel the electric force due to residual potential difference,
the total measured force was represented in the form
\begin{equation}
F_{\rm tot}(d)=F(d)-\pi\epsilon_0 R\frac{V_{\rm rms}^2}{d}-a.
\label{eq18}
\end{equation}

As a next step, Ref.~\refcite{19} performed the $\chi^2$ fit of the
mean data for the total force measured at 21 separation distances to
Eq.~(\ref{eq18}) with the two fitting parameters $V_{\rm rms}$ and $a$.
Note that the experimental errors indicated\cite{19} in the mean total
forces do not include the systematic constituents and are unreasonably
small. For example, at the largest separation, $d=7.29\,\mu$m,
the measured mean total force is equal to $F_{\rm tot}=(19.54\pm 0.28)\,$pN,
i.e., the relative error is equal to only 1.4\% (see Ref.~\refcite{38} for
a more detailed discussion of the errors in this experiment).
Here we should only stress that inclusion of the systematic errors
into the fitting procedure seems necessary.

When the Drude model approach to the Casimir force is used, from the fit
performed it was obtained\cite{19} that $\chi_{\min}^2=19.76$ with the
fitting parameters $a=-3.0\,$pN and $V_{\rm rms}=5.4\,$mV.
This was characterized as an excellent agreement with the data.
Taking into account, however, that the number of degrees of freedom
$f=19$ one obtains that $P(\chi^2>\chi_{\min}^2)=0.41$.
Such a value could be considered as being in favor of the model used
if the results of an individual measurement were fitted.
Here, however, the mean measured total force averaged over a large number
of repetitions was used in the fit. In that case the $\chi^2$-probability
should be larger than at least 50\% in order to measured data could be
considered as supporting the theoretcal model.
The plasma model approach was excluded\cite{19} basing on the following
results of the fit: $\chi_{\min}^2=608$ and $V_{\rm rms}=5.4\,$mV.

There is, however, one more weak point in the results of Ref.~\refcite{19}.
The problem is that surfaces of lenses with centimeter-size radii of
curvature have local deviations from perfect sphericity such as bubbles,
pits and scratches. The impact of such deviations on the Casimir force
was examined in detail.\cite{39,40} Specifically, it was shown that if
a surface defect is located near the point of closest approach to the
plate the simplest formulation of the PFA in Eq.~(\ref{eq5}) used in
Ref.~\refcite{19} is not applicable and should be replaced by more
sophisticated equations. The characteristic lateral sizes of invariably
present surface defects on mechanically polished and ground glass
surfaces allowed by the optical surface specification data may vary
from a few micrometers to a millimeter. In so doing their sizes in the
normal direction to the surface are below a fraction of micrometer and
their local radii of curvature may differ by tens of percent from the
lens radius $R$.\cite{39,40} This is not in contradiction with the fact
that the value of $R=15.6\,$cm was measured with the interferometric
microscope and found to vary by less than 2\% over the surface of
the lens.\cite{19}
The point is that interferometric microscopy does not provide
values of the lens radius of curvature at separate points,
but averaged values
over about 0.5\,mm regions. It was shown\cite{39,40} that at $d<3\,\mu$m
local surface defects could contribute significantly in the analysis and
simulate differences between the predictions of the Drude and plasma model
approaches.

Because of this it was suggested\cite{38} temporarily disregard all the
experimental data\cite{19} at $d<3\,\mu$m and repeat the analysis at
$d>3\,\mu$m where the influence of surface defects on the Casimir force
is negligibly small. In this case the best agreement between the data
and Eq.~(\ref{eq18}) computed using the Drude model approach is
achieved with
$\chi_{\min}^2=6.6$, $a=-0.29\,$pN and $V_{\rm rms}=5.45\,$mV.
Taking into account that $f=4$, we arrive at $P(\chi^2>\chi_{\min}^2)=0.16$.
This demonstrates a poor agreement of the data with the Drude model.

In the case of the plasma model approach the best agreement between the data
and Eq.~(\ref{eq18}) is
achieved with
$\chi_{\min}^2=2.68$, $a=3.6\,$pN and $V_{\rm rms}=4.5\,$mV.
With $f=4$ this results in $P(\chi^2>\chi_{\min}^2)=0.67$,
i.e. the plasma model approach is in
a very good agreement with the data at $d>3\,\mu$m.
This is illustrated in Fig.~6 where the experimental data for the
magnitudes of the total force multiplied by $d$ are shown as crosses
and the predictions of the plasma and Drude model approaches are given
as the black and grey lines, respectively.
\begin{figure*}[t]
\vspace*{-9.cm}
\centerline{\hspace*{3.cm}\psfig{file=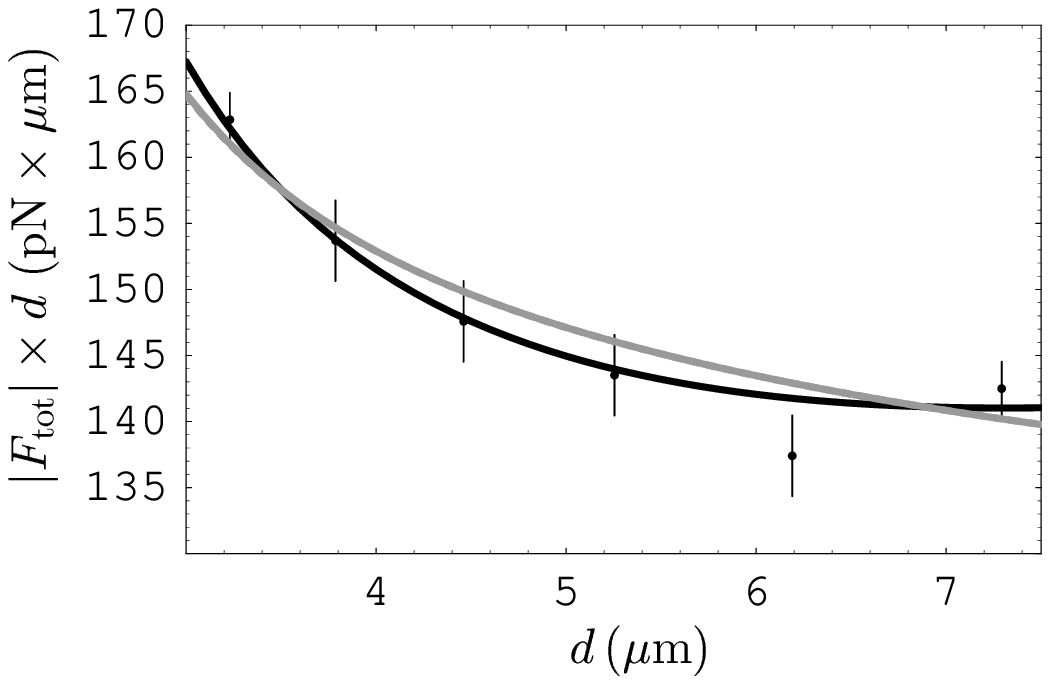,width=12cm}}
\vspace*{-4.cm}
\caption{The experimental data for the magnitudes of the mean
measured force  multiplied by
separation are shown as crosses. The arms of the crosses
indicate the experimental errors.
The grey and black lines demonstrate the best fit
 of the total theoretical force to the
experimental data
 computed using the Drude and plasma models,
respectively, with two fitting parameters.
}
\end{figure*}
\section{Conclusions}

{}From the foregoing one arrives at the following conclusions.

The comparison of the PFA with the exact results for a sphere above
a plate show that the PFA is well applicable to all performed
experiments in sphere-plate geometry.

A number of experiments demonstrate that the inclusion of relaxation
of free charge carriers for metals and dc conductivity for dielectrics
in the Lifshitz theory leads to contradictions with the measurement
data. This fact awaits for a complete theoretical explanation.

The comparison between a fitting procedure and an independent
theoretical calculation where it is available shows that the fit by
itself cannot be used to make a choice between two competing
theoretical approaches.

The suggested quasi-local model of patches is incapable to explain
the difference between the results of experiment using a micromachined
oscillator and the Drude model approach to the Casimir force.

The claimed observation of the thermal Casimir force, as predicted by the
Drude model approach, is not an independent measurement, but a fit using
two fitting parameters. At separations above 3$\,\mu$m the data of this
experiment are in agreement not with the Drude, but with the plasma
model. Below $3\,\mu$m a seeming agreement with the Drude model can be
explained by disregard of surface imperfections.

\section*{Acknowledgments}
G.L.K.\ and V.M.M.\ were partially supported by
the NSF Grant No.~PHY0970161
 and by DFG grant BO\,\,1112/20--1.


\end{document}